\documentstyle[12pt,epsf]{article}
\newcommand{\be}{\begin{equation}}
\newcommand{\ee}{\end{equation}}
\newcommand{\bea}{\begin{eqnarray}}
\newcommand{\eea}{\end{eqnarray}}
\newcommand{\nn}{\nonumber}

\begin{document}

\def\gamh{\Gamma_H}
\def\esp #1{e^{\displaystyle{#1}}}
\def\de{\partial}
\def\eb{E_{\rm beam}}
\def\deb{\Delta E_{\rm beam}}
\def\sigm{\sigma_M}
\def\sigmmax{\sigma_M^{\rm max}}
\def\sigmmin{\sigma_M^{\rm min}}
\def\sige{\sigma_E}
\def\dsigm{\Delta\sigma_M}
\def\mh{M_H}
\def\lyear{L_{\rm year}}

\def\wstar{W^\star}
\def\zstar{Z^\star}
\def\ie{{\it i.e.}}
\def\etal{{\it et al.}}
\def\eg{{\it e.g.}}
\def\pzero{P^0}
\def\mt{m_t}
\def\mpzero{M_{\pzero}}
\def\mev{~{\rm MeV}}
\def\gev{~{\rm GeV}}
\def\gam{\gamma}
\def\lsim{\mathrel{\raise.3ex\hbox{$<$\kern-.75em\lower1ex\hbox{$\sim$}}}}
\def\gsim{\mathrel{\raise.3ex\hbox{$>$\kern-.75em\lower1ex\hbox{$\sim$}}}}
\def\ntc{N_{TC}}
\def\epem{e^+e^-}
\def\tauptaum{\tau^+\tau^-}
\def\lplm{\ell^+\ell^-}
\def\anti{\overline}
\def\mw{M_W}
\def\mz{M_Z}
\def\fbi{~{\rm fb}^{-1}}
\def\mupmum{\mu^+\mu^-}
\def\rts{\sqrt s}
\def\sigrts{\sigma_{\tiny\rts}^{}}
\def\sigrtssq{\sigma_{\tiny\rts}^2}
\def\sigrtsprime{\sigma_{E}}
\def\nsigrts{n_{\sigrts}}
\def\gampzero{\Gamma_{\pzero}}
\def\pzerop{P^{0\,\prime}}
\def\mpzerop{M_{\pzerop}}

\font\fortssbx=cmssbx10 scaled \magstep2
%\hbox to \hsize{
%
%\special{psfile=uwlogo.ps
% hscale=8000 vscale=8000 hoffset=-12 voffset=-2}
%\hskip.5in \raise.1in
%
%$\vcenter{ $\hbox{\fortssbx University of Florence}
%$\hbox{\fortssbx University of Geneva} }
%
%\hfill
%
%$\vcenter{
%\hbox{\bf DFF-xxx/9/99}
%\hbox{\bf UGVA-DPT-1999 09-xxx}
%}$}

%
\medskip
\begin{center}
{\Large\bf\boldmath The Color-Flavor Locking Phase at $T \not =0
$.}\\
\rm
\vskip1pc
{\Large R. Casalbuoni$^{a,b}$ and R. Gatto$^c$}\\
%\\ D. Dominici$^{a,b}$ S. De Curtis$^b$ and R. Gatto$^c$\\}
\vspace{5mm}
{\it{$^a$Dipartimento di Fisica, Universit\`a di Firenze, I-50125
Firenze, Italia
\\
$^b$I.N.F.N., Sezione di Firenze, I-50125 Firenze, Italia\\
$^c$D\'epart. de Physique Th\'eorique, Universit\'e de Gen\`eve,
CH-1211 Gen\`eve 4, Suisse}}
\end{center}
\bigskip
\begin{abstract}

We study the color-flavor locked phase of QCD with three massless
quarks at high chemical potential and small non zero
temperatures. We make use of the recently introduced effective
action to describe such a phase. We obtain the exact order $T^2$
behaviour of the condensates and of the pressure by formally
comparing the derivative from the QCD functional with symmetry
breaking to that from the effective lagrangian with external
sources, respecting the residual $Z_2$ invariance of the
color-flavor locked phase. From these exact results, but now at a very
tentative
level of conjecture, we are lead to think that the phase structure of QCD at
very high density consists of two superconducting phases and a symmetric one.

\noindent

\end{abstract}

\section{Introduction}

QCD at high density has attracted a great interest lately.  A
recent review by Frank Wilczek \cite{wilczek0} is entitled "The
recent excitement in high-density QCD". We refer to this article
and also to a latest review by Krishna Rajagopal \cite{rajagopal}
for clear and authoritative descriptions of the subject. Numerous
authors have recently contributed to this rapidly advancing field
\cite{ba}.

A particularly interesting situation occurs in QCD with three massless
flavors at high values of the baryonic chemical potential, where
the suggested condensation pattern would imply the phenomenon of color-flavor
locking (CFL), as discussed in the papers by Alford,  Rajagopal and F.Wilczek
\cite{wilczek}, by T. Sch\"afer and F. Wilczek \cite{wilczek1}, and
by M. Alford, J. Berges, and K. Rajagopal \cite{alford}.

We have recently constructed the effective action describing color-flavor
locking \cite{CG}. In this work we use the effective action to derive some
rigorous results on the condensate behaviour for small non vanishing
temperature in the CFL phase. This we do by introducing external sources and
correlating the generating functional in QCD to the generating functional
for the goldstones. In doing this the $Z_2$ invariance left in the CFL phase
is taken into account. We add to our exact results some tentative conjectural
considerations on the phase structure of QCD at high density.

\section{The effective theory}
In this Section we shortly review the effective action for the
color-flavor locked (CFL) phase of QCD introduced in
\cite{CG}. The CFL phase is characterized by the symmetry
breaking pattern \cite{wilczek,wilczek1,alford}
\be
G=SU(3)_c\otimes SU(3)_L\otimes SU(3)_R \otimes U(1)\to
H=SU(3)_{c+L+R}
\ee
from dynamical formation of condensates of the type
\be
\langle\psi_{ai}^L\psi_{bj}^L\rangle=-\langle\psi_{ai}^R\psi_{bj}^R
\rangle=\gamma_1\delta_{ai}\delta_{bj}+\gamma_2\delta_{aj}\delta_{bi}
\label{2}
\ee
where $\psi_{ai}^{L(R)}$ are Weyl  spinors and a sum over spinor
indices is understood. The indices $a,b$ and $i,j$ refer to
$SU(3)_c$ and to $SU(3)_L$ (or $SU(3)_R$) respectively. In fact $H$ contains
an additional factor
$Z_2$ which will play an essential role in the following.

The effective lagrangian describes the CFL phase for
momenta smaller than the energy gap (the existing numerical
estimates range between 10 and 100 $MeV$). The relevant degrees
of freedom are the massless goldstones arising from the
symmetry breaking. Before the gauging of $SU(3)_c$  we need 17
goldstones. We introduce coset matrix
fields $X$ and $Y$ transforming under $G$  as a
left-handed and a right-handed quark field respectively. We require
\be
X\to g_c X g_L^T,~~~~~Y\to g_c Y g_R^T
\ee
with $g_c\in SU(3)_c,~~~g_L\in SU(3)_L,~~~g_R\in SU(3)_R$.
$X$ and $Y$ are $SU(3)$ matrices, breaking respectively
$SU(3)_c\otimes SU(3)_L$ and $SU(3)_c\otimes SU(3)_R$.
This gives 16
goldstones. The additional one describes the
breaking of the baryon number. This is done in terms of a $U(1)$
factor transforming under $G$ as
\be
U\to g_{U(1)} U,~~~~g_{U(1)}\in U(1)
\ee
We define the following anti-hermitian and traceless
currents
\be
J_X^\mu=X\de^\mu X^\dagger,~~~ J_Y^\mu=Y\de^\mu Y^\dagger,~~~
J_\phi=U\de^\mu U^\dagger
\ee
which transform under
the global group $G$ as
\be
J_X^\mu\to g_c J_X^\mu g_c^\dagger,~~~J_Y^\mu\to g_c J_Y^\mu
g_c^\dagger,~~~J_\phi^\mu\to J_\phi^\mu
\ee
 Barring  WZW terms \cite{wess} (see ref. \cite{CG} for a brief
discussion), the most general rotational symmetric (at finite
density Lorentz  invariance is broken to $O(3)$) lagrangian
invariant under $G$, with at most two derivatives, is given by
\bea
{\cal L}&=&-\frac{F_T^2}4 Tr[(J_X^0-J_Y^0)^2]-
\alpha_T\frac{F_T^2}4 Tr[(J_X^0+J_Y^0)^2] -\frac{f_T^2}2 (J_\phi^0)^2\nn\\
&+&\frac{F_S^2}4 Tr[(\vec J_X-\vec J_Y)^2]+
\alpha_S\frac{F_S^2}4 Tr[(\vec J_X+\vec J_Y)^2] +\frac{f_S^2}2 (\vec J_\phi)^2
\label{BESS}
\eea
We have required  invariance under parity, that is symmetry under
$X\leftrightarrow Y$.

With the parametrization
\be
X=\esp{i\tilde\Pi_X^aT_a},~~~
Y=\esp{i\tilde\Pi_Y^aT_a},~~~ U=\esp{i\tilde\phi},~~~
a=1,\cdots 8
\ee
(where the  $SU(3)$ matrices $T_a$  satisfy
$Tr[T_aT_b]=\frac 1 2 \delta_{ab}$)
and by defining
\be
\Pi_X=\sqrt{\alpha_T}\,\frac{F_T}2(\tilde\Pi_X+\tilde\Pi_Y),~~~
\Pi_Y=\frac{F_T} 2(\tilde\Pi_X-\tilde\Pi_Y),~~~
\phi=f_T\tilde\phi
\label{rescaling}
\ee
 the  kinetic term is
\be
{\cal L}_{\rm kin}=\frac 12 ({\dot\Pi_X}^a)^2+
\frac 12
({\dot\Pi_Y}^{a})^2+ \frac 1 2 (\dot\phi)^2
-\frac {v_X^2}2 |\vec\nabla\Pi_X^{a}|^2-
\frac {v_Y^2}2
|\vec\nabla\Pi_Y^{a}|^2- \frac {v_\phi^2} 2 |\vec\nabla\phi|^2
\ee
where
\be
v_X^2=\frac{\alpha_S F_S^2}{\alpha_T F_T^2},~~~v_Y^2=\frac{F_S^2}
{F_T^2},
~~~v_\phi^2=\frac{f_S^2}{f_T^2}
\label{velocity}
\ee
The three different types of goldstones  move
with different velocities, still satisfying a linear
dispersion relation $E=vp$.

The lagrangian of eq. (\ref{BESS}) must have a local $SU(3)_c$
invariance inherited by the color group of QCD. To make
it explicit we only need to substitute  usual
derivatives with  covariant ones
\be
\de_\mu X\to D_\mu X= \de_\mu X-g_\mu X,~~~
\de_\mu Y\to D_\mu Y= \de_\mu Y-g_\mu Y,~~~ g_\mu\in {\rm Lie}~SU(3)_c
\ee
The  currents become
\be
J_X^\mu=X\de^\mu X^\dagger+g^\mu,~~~ J_Y^\mu=Y\de^\mu
Y^\dagger+g^\mu
\ee
giving the lagrangian
\bea
{\cal L}&=&-\frac{F_T^2} 4Tr[(X\de^0 X^\dagger -Y\de^0
Y^\dagger)^2]-\alpha_T\frac{F_T^2} 4Tr[(X\de^0 X^\dagger +Y\de^0
Y^\dagger+2g^0)^2]\nn\\ &-&\frac{f_T^2}2(J_\phi^0)^2+{\rm
spatial~
 terms~and~kinetic~part~for}~ g^\mu
\label{lagr BESS}
\eea
We define
\be
g_\mu=ig_s\frac{T_a}2 g_\mu^a
\ee
where $g_s$  is the QCD coupling constant. The gluon
field acquires a mass. This can easily be seen in a gauge such
that $X=Y^\dagger$. This implies
\be
\tilde\Pi_X=-\tilde\Pi_Y
\ee
or
\be
\Pi_X=0,~~~\Pi_Y=F_T\tilde\Pi_X
\ee
This gauge is the unitary one (the bilinear term in the
goldstones and in the gluon field in eq. (\ref{lagr BESS}) is
proportional to $g^\mu\de_\mu(\tilde\Pi_X+\tilde\Pi_Y)$ and
cancels out). The gluon mass (for the expected velocities of
order one) is given by
\be
m_g^2=\alpha_T g_s^2\frac {F_T^2}4
\ee
The $X\leftrightarrow Y$ symmetry, in this
gauge, implies $\Pi_Y\leftrightarrow -\Pi_Y$.

For goldstone energies much smaller than the gluon mass we can
neglect the gluon kinetic term. The lagrangian (\ref{lagr BESS})
is then the hidden gauge symmetry version of the chiral
lagrangian for QCD \cite{BESS} (except for the contribution of
the field $\phi$). In fact, in this limit, the gluon field
becomes an auxiliary field which can be eliminated through its
equation of motion
\be
g_\mu=-\frac 1 2(X\de_\mu X^\dagger+Y\de_\mu Y^\dagger)
\ee
obtaining
\be
{\cal L}=-\frac{F_T^2}4Tr[(X\de^0 X^\dagger-Y\de^0
Y^\dagger)^2]-\frac{f_T^2} 2(J_\phi^0)^2+ {\rm spatial~terms}
\label{QCD}
\ee
or
\be
{\cal L}=\frac{F_T^2}4 \left(Tr[\dot \Sigma\dot \Sigma^\dagger]-
v_Y^2 tr[\vec\nabla \Sigma\cdot\vec\nabla \Sigma^\dagger]\right)
-\frac{f_T^2}2\left((J_\phi^0)^2-v_\phi^2|\vec J_\phi|^2\right)
\ee
where $\Sigma=Y^\dagger X$ transforms under the group $SU(3)_c\otimes
SU(3)_L\otimes
SU(3)_R$ as $\Sigma\to g_R^* \Sigma g_L^T$. The goldstone $\phi$
 could be interpreted, according to ref. \cite{wilczek1}, as a dibaryon state,
 $(udsuds)$, particularly light, as had been
pointed out by R. Jaffe \cite{jaffe} (we had called it the "deus ex
machina" state). In total, after the breaking of the color group one is left
with the massless photon and 9 physical Goldstone bosons
transforming as the representation $1+8$ of the unbroken SU(3).

\section{Introduction of $Z_2$-preserving breaking and of external sources}

We shall introduce external sources \cite{gasser} to correlate
the QCD generating functional and the generating functional for
the goldstones. We consider goldstone momenta smaller than the
Fermi momentum (or the chemical potential $\mu$). We recall that,
although the goldstones do not necessarily move at the speed of
light, they satisfy a linear dispersion relation
\be
E=v|\vec p|
\ee
where $v$ is the appropriate velocity (see eq. (\ref{velocity})).
We will need  a relation between the
condensates and the parameters appearing in the effective lagrangian.
 To this end we
introduce in the QCD lagrangian a breaking term with the same
structure of the condensates, that is
\be
{\cal L}_{QCD}^{\rm break}={\cal
M}^L_{aibj}\psi_{ai}^L\psi_{bj}^L- {\cal
M}^R_{aibj}\psi_{ai}^R\psi_{bj}^R+c.c.
\label{32}
\ee
where $\psi_{ai}^{L(R)}$ are Weyl  spinors and a sum over spinor
indices is understood. The indices $a,b$ and $i,j$ refer to
$SU(3)_c$ and to $SU(3)_L$ (or $SU(3)_R$ respectively).  ${\cal
M}_{aibj}(x)$ is a set of external scalar fields.  At the same time we
  introduce a breaking term in the effective lagrangian
\bea
{\cal L}_{\rm eff}^{\rm break}&=&\left(\gamma_1{\cal M}^L_{aibj}+
\gamma_2{\cal M}^L_{ajbi}\right ) X_{ai}X_{bj}U^2\nn\\&+&
\left(\gamma_1{\cal M}^R_{aibj}+
\gamma_2{\cal M}^R_{ajbi}\right) Y_{ai}Y_{bj}U^2+c.c.
\eea
where the $U^2$ factor takes into account  that the term in eq.
(\ref{32}) breaks the $U(1)_V$ in such a way to preserve a $Z_2$
symmetry. To get the desired relation  one has simply to
differentiate the  QCD generating functional  and that derived
from the effective lagrangian and equate the results. Since in
the vacuum one has
\be
\langle X_{ia}\rangle=\langle Y_{ia}\rangle=\delta_{ia},~~~
\langle U\rangle =1
\ee
 the contribution to ${\cal L}_{\rm eff}$ is given by
\be
{\cal L}_{\rm eff}^{\rm break}\to (\gamma_1{\cal M}^L_{aibj}+
\gamma_2{\cal M}^L_{ajbi}) \delta_{ai}\delta_{bj}+
(\gamma_1{\cal M}^R_{aibj}+
\gamma_2{\cal M}^R_{ajbi}) \delta_{ai}\delta_{bj}+c.c.
\ee
from which
\bea
\langle\psi_{ai}^L\psi_{bj}^L\rangle&=&\gamma_1\delta_{ai}\delta_{bj}
+\gamma_2\delta_{aj}\delta_{bi}\nn\\
\langle\psi_{ai}^R\psi_{bj}^R\rangle&=&-\gamma_1\delta_{ai}\delta_{bj}
-\gamma_2\delta_{aj}\delta_{bi}
\label{condensates}
\eea
Now let us consider the contribution to the goldstones masses of a
breaking term corresponding to the following choice for the
external scalar source
\be
{\bar{\cal M}}^L_{aibj}={\bar{\cal M}}^R_{aibj}=
m_1\delta_{ai}\delta_{bj}+m_2\delta_{aj}\delta_{bi}
\label{sources}
\ee
with $m_1$ and $m_2$ real parameters. The projection of the
condensates along these directions gives
\bea
{\bar{\cal M}}^L_{aibj}\langle\psi_{ai}^L \psi_{bj}^L\rangle&=&
m_1 u_1^L+m_2 u_2^L\nn\\ {\bar{\cal
M}}^R_{aibj}\langle\psi_{ai}^R
\psi_{bj}^R\rangle&=& m_1 u_1^R+m_2 u_2^R
\eea
where
\be
u_1^{L,R}=\langle\psi_{aa}^{L.R}\psi_{bb}^{L,R}\rangle,~~~
u_2^{L.R}=\langle\psi_{ab}^{L,R}L\psi_{ba}^{L,R}\rangle
\ee
and using the expression (\ref{condensates}) we get
\be
u_1^{L,R}= \pm (N^2\gamma_1+N\gamma_2),~~~u_2^{L,R}=\pm
(N^2\gamma_2+N\gamma_1)
\label{35}
\ee
(for possible future uses we do this calculation for  color
and flavor groups $SU(N)$). These expressions can be easily
inverted obtaining
\be
\gamma_1=\frac{Nu_1-u_2}{N(N^2-1)},~~~
\gamma_2=\frac{Nu_2-u_1}{N(N^2-1)}
\label{relations}
\ee
where
\be
u_{i}=u_{i}^L=-u_{i}^R,~~~~i=1,2
\ee
The breaking term in the effective lagrangian, taken at the
values (\ref{sources}) for the external sources, gives rise to
mass terms for the goldstones . In fact, from
\bea
{\cal L}_{\rm eff}^{\rm masses}&=& (\gamma_1{\bar{\cal
M}}_{aibj}^L+ \gamma_2{\bar{\cal M}}_{ajbi}^L)X_{ai}X_{bj}U^2+
(X\leftrightarrow Y) +c.c.\nn\\&=& (m_1\gamma_1+m_2\gamma_2)
(Tr[X])^2U^2+ (m_1\gamma_2+m_2\gamma_1)Tr[X^2]U^2\nn\\\ &+&
(X\leftrightarrow Y) +c.c.
\eea
evaluating the traces at the second order in the Goldstone
fields, we get
\be
(tr[X])^2\approx(N-\frac 1 4 \tilde\Pi_X^2)^2= N^2-\frac N
2\tilde\Pi_X^2
\ee
\be
Tr[X^2]\approx N-\tilde\Pi_X^2
\ee
and analogous expressions for the $Y$ fields. Therefore
\bea
{\cal L}_{\rm eff}^{\rm masses}&=&4\left(m_1(N^2\gamma_1+N\gamma_2)+
m_2(N^2\gamma_2+N\gamma_1)\right)\nn\\
&-&8\left(m_1(N^2\gamma_1+N\gamma_2)+m_2(N^2\gamma_2+N\gamma_1)\right)
\tilde\phi^2\nn\\
&-&\left((N\gamma_1+2\gamma_2)m_1+
(2\gamma_1+N\gamma_2)\right)\left(\tilde\Pi_X^2+\tilde\Pi_Y^2\right)
\eea
Using
\be
\tilde\Pi_X^2+\tilde\Pi_Y^2=\frac 2 {\alpha_T F_T}\Pi_X^2+\frac 2{F_T}\Pi_Y^2
\ee
and eqs. (\ref{35}) and (\ref{relations}) we get
\bea
&&{\cal L}_{\rm eff}^{\rm masses}=
4(m_1u_1+m_2u_2)-8\frac{m_1u_1+m_2u_2} {f_T^2}\phi^2\nn\\
&-&2\frac{((N^2-2)u_1+Nu_2)m_1+((N^2-2)u_2+Nu_1)m_2} {\alpha_T
F_T^2 N^2 (N-1)}\Pi_X^2\nn\\&-&
2\frac{((N^2-2)u_1+Nu_2)m_1+((N^2-2)u_2+Nu_1)m_2} {F_T^2 N^2
(N-1)}\Pi_Y^2
\eea
Therefore the goldstones masses (that is twice the coefficient of
the square terms in the Goldstone fields) are given by
\bea
m_\phi^2&=& \frac  {16}{f_T^2}(m_1u_1+m_2u_2)\nn\\ m_X^2&=&4
\frac{((N^2-2)u_1+Nu_2)m_1+((N^2-2)u_2+Nu_1)m_2} {\alpha_T F_T^2 N^2
(N-1)}\nn\\
m_Y^2&=&4\frac{((N^2-2)u_1+Nu_2)m_1+((N^2-2)u_2+Nu_1)m_2} {F_T^2
N^2 (N-1)}
\eea

\section{Exact behaviour of the condensates and of the pressure
to the order $T^2$}

We can now evaluate the behaviour of the condensates in the low
temperature limit up to the order $T^2$. For the corresponding case of
chiral symmetry breaking at zero densities the authors of ref.
\cite{gasser} had shown that, in the infinite volume
limit and for small pion masses, it is sufficient at small temperatures
 to take into account
only the contribution from free pions. The same reasonings go on in our case.
Thus, by writing the
partition function in the form
\be
Tr[\esp{-\beta H}]=\esp{-\beta L^3 z}
\ee
we get, for a single free goldstone of mass $M$
\be
z=\epsilon_0-\frac 12 g_0(M^2,T)
\ee
where $\epsilon_0$ is the energy density of the ground state and
\be
g_0(M^2,T)=2T\int\frac{d^3\vec p}
{2\pi^3}\left[-\ln\left(1-\exp(-E/T)\right)\right]
\ee
with
\be
E=\sqrt{v^2|\vec p|^2+M^2}
\ee
The previous expression can be reduced to the standard expression
through the change of variables $\vec p\,'=v\vec p$, obtaining
\be
g_0(M^2,T)=\frac{2T}{v^3}\int\frac{d^3\vec
p'}{2\pi^3}\left[-\ln\left(1-\exp(-E'/T)\right)
\right]
\ee
with
\be
E'=\sqrt{|\vec p\,|^2+M^2}
\ee
In the limit $M^2\to 0$ we get
\be
g_0=\frac{1}{v^3}\frac{\pi^2 T^4}{45}
\ee
and
\be
-\frac{\de g_0}{\de M^2}=\frac{1}{v^3}\frac{T^2}{12}
\ee
We can now derive the expression for the pressure in the massless limit
\be
P=\frac 1 2(N^2-1) g_0(m_Y^2,T) +\frac 1 2 g_0(m_\phi^2,T)
\ee
and, for $v_Y\approx v_\phi\approx 1$ and $T\to 0$ we get the simple
result
\be
P\approx \frac{\pi^2}{90}N^2T^4
\ee
The condensates can be easily evaluated in terms of the energy
density $z$
\be
u_i(T)=-\frac 1 {4}\frac{\de}{\de m_i} z
\ee
as it follows from eq. (\ref{32}) for the chosen external scalar
sources.  In our case we have
\be
z=-4(m_1 u_1+m_2 u_2)-\frac 1 2 (N^2-1) g_0(m_Y^2,T)-
\frac 1 2 g_0(m_\phi^2,T)
\ee
Therefore
\bea
u_1(T)&=&u_1-(N^2-1)\frac {(N^2-2) u_1+Nu_2}{N^2(N-1)}
\frac{T^2}{24 F_T^2v_Y^3}-u_1\frac {T^2}{6 f_T^2 v_\phi^3}\nn\\
u_2(T)&=&u_2-(N^2-1)\frac {(N^2-2) u_2+Nu_1}{N^2(N-1)}
\frac{T^2}{24 F_T^2v_Y^3}- u_2\frac {T^2}{6 f_T^2 v_\phi^3}
\label{54}
\eea

More complicate
calculations along the lines of ref. \cite{gasser} would  allow to calculate
an additional term in the expansion in $T^2$.

In the case of the analogous calculation made for the chiral
phase one gets an insight about the possibility of a second order
phase transition simply by noticing that the chiral condensate
(evaluated for small T) decreases with the temperature. Of
course, this conjecture should be supported by higher order
calculations, or even better by a complete calculation at any
order in the temperature, but it looks as a reasonable situation.
The question is, if in the present situation one can get a similar
insight. To this end, we begin by the observation that now the
condensate is determined by two quantities, $\gamma_1$ and
$\gamma_2$ or their combinations $u_1$ and $u_2$. Speaking in
terms of $(u_1,u_2)$, one can look at these two variables as the
components of a vector in a two-dimensional space.  To look for a
signal of a phase transition one should be able to decide if the
length of the vector decreases or increases with the temperature.
This can be done by evaluating directly the length of the vector
from eq. (\ref{54}), or better by using a reference frame where
the vector does not rotate (by varying $T$). This reference frame
is simply obtained by the following change of coordinates
\be
v_1=u_1+u_2,~~~v_2=u_1-u_2
\ee
Then, from eq. (\ref{54}) we get
\bea
v_1(T)&=&v_1(0)\left[1-\frac {(N^2+N-2)(N+1)}{N^2}
\frac{T^2}{24 F_T^2v_Y^3}-\frac {T^2}{6 f_T^2v_\phi^3}\right]\nn\\
v_2(T)&=&v_2(0)\left[1-\frac {(N^2-N-2)(N+1)}{N^2}
\frac{T^2}{24 F_T^2v_Y^3}- \frac {T^2}{6 f_T^2 v_\phi^3}\right]
\label{55}
\eea
showing that around zero temperature both components decrease for
increasing $T$.

For $N=3$ and all the couplings of the same order of magnitude
one gets
\be
\frac{v_1(T)}{v_1(0)}\approx
1-\frac{19}{54}\frac{T^2}{F^2},~~~\frac{v_2(T)}{v_2(0)} \approx
1-\frac{13}{54}\frac{T^2}{F^2}
\ee
where $F$ is the common value of the couplings.

So far the results are exact (except for the assumptions of equal
coupling in the last formulae) to the order $T^2$. Extrapolating
from these results may be very dangerous as we do not know
anything on the full expressions. However one can try to imagine
the possible scenarios coming from a naive extrapolation of the
previous results.

An interesting possibility,  suggested by the eqs. (\ref{55}), is that there
exist two different temperatures $T_1$ and $T_2$, $T_2\ge T_1$, such that
$v_1(T)=0$ for $T\ge T_1$ and $v_2(T)=0$ for $T\ge T_2$ (we will
discuss later the opposite possibility). Since $v_1=0$ means $u_1=-u_2$, and
\be
\gamma_1=\frac N{N-1} u_1=-\gamma_2
\ee
we see that the condensate of eq. (\ref{2}) becomes antisymmetric
in the exchange of the color (or of the flavor) indices. In the
case $N=3$ one can write
\be
\langle\psi_{ai}^L\psi_{bj}^L\rangle=-\langle\psi_{ai}^R\psi_{bj}^R
\rangle=\gamma_1\epsilon_{abm}\epsilon_{ijm}
\label{Cooper}
\ee
The possibility that the condensate at $T=0$ in the CFL phase may
assume this form was considered in ref. \cite{wilczek}, with the
conclusion that this is disfavored by the dynamics. But here we
see that it is possible that this situation can be realized at
$T\not =0$. The equality $\gamma_1=-\gamma_2$ has a clear
group-theoretical meaning since the condensate (\ref{2}) behaves,
under $SU(3)_c\otimes SU(3)_{L,R}$ as $(3,3)\otimes (6,6)$ due to
the symmetry under the overall exchange of the color and flavor
indices. It is easy to show that the combination
$\gamma_1-\gamma_2$ corresponds to the representation $(3,3)$,
whereas $\gamma_1+\gamma_2$ to the representation $(6,6)$.
Therefore, $\gamma_1=-\gamma_2$ means that only the component
$(3,3)$ survives. On the other hand, from the point of view of
symmetry breaking the situation at $T_1\le T\le T_2$ would not be
very much different than at lower temperatures, since still we
have the same general type of symmetry breaking. In our
hypothesis the symmetry would be restored only at $T\ge T_2$,
when both $\gamma_1$ and $\gamma_2$ vanish.

We see that, under this  assumption, there would be two
superconducting phases differing only in the symmetry of the
Copper pairs wave function. In one phase (say SC1 at $T\le T_1$)
the wave function is symmetric in the simultaneous exchange of
the color and flavor indices. In the other phase (SC2 at $T_1\le
T\le T_2$), the wave function is antisymmetric in the separate
exchange of color and flavor indices. It is interesting to notice
that the wave function in the phase SC2 is similar to the one
obtained in the case of two flavors (see ref. \cite{wilczek}).

Our exact results may thus lead to conjecture, after the above
extrapolation, that at high density the phase structure of QCD is
given by two different superconducting phases and by a symmetric
phase. Increasing the temperature one goes from the SC1 phase
through a sort of crossover to the SC2 phase and after that the
symmetry is restored.

Although our  extrapolation from the behaviour close to $T=0$
gives indications that $v_1$ goes to zero before $v_2$, one can
ask what would happen in the opposite situation. We see
immediately that when $v_2=0$ one has $u_1=u_2$ and
\be
\gamma_1=\frac 1{N(N-1)} u_1=-\gamma_2
\ee
Therefore, also in this case one obtains for the condensate the
expression in eq. (\ref{Cooper}).

We stress again that, differently from the results at order
$T^2$,
 which are the main object of this note,
the considerations we have just added for larger $T$ are highly speculative
and go much beyond the exact results we have derived.

\section{Conclusions}
We have used the effective action for the color-flavor locked
phase of QCD with three massless quarks at high chemical
potential and derived exact behaviours at order $T^2$ by formally
introducing a QCD symmetry breaking term of the same structure of
the condensates and a corresponding term in the effective
lagrangian, and by equating the derivative from the QCD
functional to that from the effective lagrangian. More complicate
calculations, not done
here, would  allow to calculate  an additional term in the
expansion in $T^2$. Although it would be unreliable to
extrapolate from the obtained low $T$ results, one may very tentatively
imagine the existence of two superconducting phases with a
crossover between them and an ensuing second order phase transition
to the symmetric phase, when increasing $T$ at
a fixed large chemical potential.


\begin{thebibliography}{99}


\bibitem{wilczek0}
F. Wilczek, invited talk at Panic 99, Uppsala, June 1999,
hep-ph/9908480

\bibitem{rajagopal}
K. Rajagopal, invited talk at Quark Matter 99, hep-ph/9908360

\bibitem{ba}
F. Barrois, Nucl. Phys. {\bf B129}, 390 (1977); S. C. Frautschi,
in Workshop on hadronic matter at extreme densities, Erice 1978;
D. Bailin and A. Love, Phys. Rep. {\bf 107}, 325 (1984); M.
Alford, K. Rajagopal and F. Wilczek, Phys.Lett. {\bf B422} (1998)
247; R. Rapp, T. Sch\"affer, E. V. Shuryak and M. Velkovsky,
Phys. Rev. Lett. {\bf 81} (1998) 53; M. Alford, K. Rajagopal and
F. Wilczek, Nucl.Phys. {\bf B537} (1999) 443; J. Berges and K.
Rajagopal, Nucl. Phys. {\bf B538}, 215 (1999); T. Sch\"affer and
F. Wilczek, Phys. Rev. Lett. {\bf 82} (1999) 3956; {\it ibidem}
hep-ph/9903503; M. Alford, J. Berges, and K. Rajagopal,
hep-ph/9903502; T. Sch\"afer and F. Wilczek, Phys. Lett. {\bf
B450}, 325 (1999); D. T. Son, Phys. Rev. D {\bf 59}, 094019
(1999); N. Evans, S. D. H. Hsu, and M. Schwetz, Nucl. Phys. {\bf
B551}, 275 (1999); G. W. Carter and D. Diakonov, Phys. Rev. D
{\bf 60}, 016004 (1999); N. Evans, S. Hsu, M. Schwetz,
hep-ph/9810514; R. D. Pisarski and D. H. Rischke,
nucl-th/9811104; K. Langfeld and M. Rho, hep-ph/9811227; J.
Berges, D. U. Jungnickel, and C. Wetterich, hep-ph/9811387; D.K.
Hong, hep-ph/9812510; N. O. Agasian, B. O. Kerbikov, and V. I.
Shevchenko, hep-ph/9902335; R. D. Pisarski and D. H. Rischke,
nucl-th/9903023; T. M. Schwarz, S. P. Klevansky, and G. Papp,
nucl-th/9903048; D. G. Caldi and A. Chodos, hep-ph/9903416; R.
Rapp, T. Sch\"afer, E. V. Shuryak, and M. Velkovsky,
hep-ph/9904353; E. Shuster and  D.T. Son, hep-ph/9905448; D. K.
Hong, hep-ph/9905523; R. D. Pisarski and D. H. Rischke,
nucl-th/9906050; D. K. Hong, V. A. Miransky, I. A. Shovkovy, and
L. C. R. Wijewardhana, hep-ph/9906478; T. Sch\"afer and F.
Wilczek, hep-ph/9906512; D. K. Hong, M. Rho and I. Zahed,
hep-ph/9906551; R. D. Pisarski and D. H. Rischke,
nucl-th/9907041; R.D.Pisarski and D.H.Rischke nucl-th/9907094; V.
A. Miransky, I. A. Shovkovy and L. C. R. Wijewardhana,
hep-ph/9908212; M. Alford, J. Berges, K. Rajagopal,
hep-ph/9908235; W. E. Brown, J. T. Liu and H. Ren,
hep-ph/9908248; E. V. Shuryak,hep-ph/9908290; S. D. H. Hsu and M.
Schwetz, hep-ph/9908310; G. W. Carter and D. Diakonov,
hep-ph/9908314; A. Chodos, F. Cooper, W. Mao, H. Minakata and A.
Singh. hep-ph/9909296.



\bibitem{wilczek}
M. Alford, K. Rajagopal and F. Wilczek, Nucl.Phys.
{\bf B537} (1999) 443.

\bibitem{wilczek1}
T. Sch\"affer and F. Wilczek, Phys. Rev. Lett. {\bf 82} (1999)
3956; {\it ibidem} hep-ph/9903503;

\bibitem{alford}
M. Alford, J. Berges, and K. Rajagopal, hep-ph/9903502.


\bibitem{CG}
R. Casalbuoni and R. Gatto, hep-ph/9908227 (to appear in
Phys.Lett.B).


\bibitem{wess}
J. Wess and B. Zumino, Phys. Lett. {\bf  B37} (1971) 95; E.
Witten, Nucl. Phys. {\bf B223} (1983) 433.


\bibitem{BESS}
A.P. Balachandran, A. Stern, and G. Trahern, Phys. Rev. {\bf D19}
(1979) 2416; M. Bando, T. Kugo and K. Yamawaki,  Phys. Rep. {\bf 164}
(1988) 217;
R. Casalbuoni et al.,  Phys. Lett. {\bf 292B}  (1992) 371.


\bibitem{jaffe}
R. Jaffe, Phys. Rev. Lett. {\bf 38} (1977) 195, 617(E).


\bibitem{gasser}
J. Gasser and H. Leutwyler, Ann. Phys. {\bf 158} (1984) 142; {\it
ibidem} Nucl. Phys. {\bf B250} (1985) 465; J. Gasser and H.
Leutwyler, Phys. Lett. {\bf B184} (1987) 83; {\it ibidem}, Phys.
Lett. {\bf B188} (1987) 477.

\end{thebibliography}
\end{document}